\documentclass[aps,prl,twocolumn,floatfix,reprint]{revtex4}
\usepackage{amsfonts}
\usepackage{mathrsfs}
\usepackage{color}
\usepackage{xspace}
\usepackage{subfigure}
\usepackage{longtable}
\usepackage{xspace}
\usepackage{multirow}
\usepackage{tabularx}
\usepackage{amsmath}
\usepackage{amssymb}
\usepackage{graphicx}
\usepackage{dcolumn}
\usepackage{bm}
\usepackage{ulem}
\usepackage{hyperref}

\begin{document}

\title{Floquet FFLO superfluids and Majorana fermions in a shaken fermionic
optical lattice}
\author{Zhen Zheng$^{1,2}$}
\thanks{These authors contributed equally to this work}
\author{Chunlei Qu$^{1}$}
\thanks{These authors contributed equally to this work}
\author{Xubo Zou$^{2}$}
\author{Chuanwei Zhang$^{1}$}
\thanks{chuanwei.zhang@utdallas.edu}

\begin{abstract}
Fulde-Ferrell-Larkin-Ovchinnikov (FFLO) superfluids, Cooper pairings with
finite momentum, and Majorana fermions (MFs), quasiparticles with
non-Abelian exchange statistics, are two topics under intensive
investigation in the past several decades, but unambiguous experimental
evidences for them have not been found yet in any physical system. Here we
show that the recent experimentally realized cold atom shaken optical
lattice provides a new pathway to realize FFLO superfluids and MFs. By
tuning shaken lattice parameters (shaking frequency and amplitude), various
coupling between the $s$- and $p$-orbitals of the lattice (denoted as the
pseudo-spins) can be generated. We show that the combination of the inverted
$s$- and $p$-band dispersions, the engineered pseudo-spin coupling, and the
attractive on-site atom interaction, naturally allows the observation of
FFLO superfluids as well as MFs in different parameter regions. While
without interaction the system is a topological insulator (TI) with edge
states, the MFs in the superfluid may be found to be in the conduction or
valence band, distinguished from previous TI-based schemes that utilize edge
states inside the band gap.
\end{abstract}

\affiliation{$^{1}$Department of Physics, The University of Texas at Dallas, Richardson,
TX, 75080 USA \\
$^{2}$Key Laboratory of Quantum Information, University of Science and
Technology of China, Hefei, Anhui, 230026, People's Republic of China}
\maketitle

Optical lattices for ultra-cold atoms provide a generic platform for quantum
simulation of various condensed matter phenomena because of their precise
control of the system parameters and the lack of disorder \cite{Bloch-NP}.
In a static optical lattice, the Bloch bands are well separated by large
energy gaps and usually only one band plays a dominate role in the static
and dynamical properties of ultra-cold atoms \cite%
{Jaksch-1998,Greiner,Bloch-RMP}. Current optical lattice study has mainly
focused on the lowest $s$-orbital band, but higher-orbital (e.g., $p$%
-orbital band, etc.) physics has also been investigated extensively in both
theory \cite{Liu-2006,Wu-2008, Zhang-2011, Xu-2013} and experiment \cite%
{Wirth-2011,Sengstock-2012} in the past decade. Recently, the experimentally
realized shaken optical lattices opens a completely new avenue for studying
the physics originating from the coupling between different orbital bands
induced by the lattice shaking \cite{Chin}. It was shown in experiment that
the hybridization of $s$-band and $p$-band of a Bose-Einstein condensate
(BEC) in a shaken lattice can cause a change of the energy dispersion from a
parabolic to a double well structure, yielding a paramagnetic to
ferromagnetic phase transition \cite{Chin,Zhai-1,Erich}. More generally, by
varying the shaking parameters, various coupling between different Bloch
bands can be engineered to implement artificial gauge fields for cold atoms,
yielding exciting new exotic physics \cite{Lim-2008,Eckardt}, such as the
recent experimental observation of topological Haldane model and the
associated anomalous Hall effect \cite{Jotzu2014}.

In this paper, we investigate new superfluid physics emerged from the
coupling between the lowest two Bloch bands ($s$- and $p$-bands) in a shaken
fermionic optical lattice. The $s$- and $p$-orbitals are denoted as two
pseudo-spins, whose energy dispersions are inverted, in contrast to the same
dispersion for usual spins. We show that such inverted band dispersions,
together with attractive on-site interaction between atoms on $s$- and $p$%
-bands, provide a natural way to realize the long-sought
Fulde-Ferrell-Larkin-Ovchinnikov (FFLO) superfluids \cite{FF64,LO64}. FFLO
states, first proposed in 1960s to describe the finite momentum Cooper
pairings with a large Zeeman field, are a central concept for understanding
many exotic phenomena in different physics branches. However, unambiguous
experimental signature of FFLO states is still lacking for any physical
system. Because of the inverted band dispersion, the FFLO state becomes the
natural ground state of the system even without an explicit external Zeeman
field.

When the shaking frequency and amplitude are tuned to certain regime, the
coupling between two pseudo-spins may depend on the lattice quasi-momentum,
analogous to the artificial spin-orbit coupling (SOC) \cite%
{Ian,Pan,Peter,Zhang,Martin}. Without interaction, the system behaves like a
topological insulator (TI) \cite{Kane,Qi} and supports topological edge
states. We show that there is a quantum phase transition from FFLO to BCS
superfluids with increasing SOC. In a large parameter regime, the BCS
superfluid is topological and supports Majorana fermions (MFs) which are
localized at the lattice boundaries~\cite{Zhang2008, Sato}. MFs are
quasiparticles that are their own antiparticles and possess non-Abelian
exchange statistics, a crucial element for topological quantum computation.
More interestingly, the topological BCS superfluids and MFs may utilize the
conduction or valence bands of such a TI, instead of the edge states inside
the band gap that are commonly used in previous TI-based schemes for MFs
\cite{Fu}.

\begin{figure}[tbp]
\centering\includegraphics[width=0.48\textwidth]{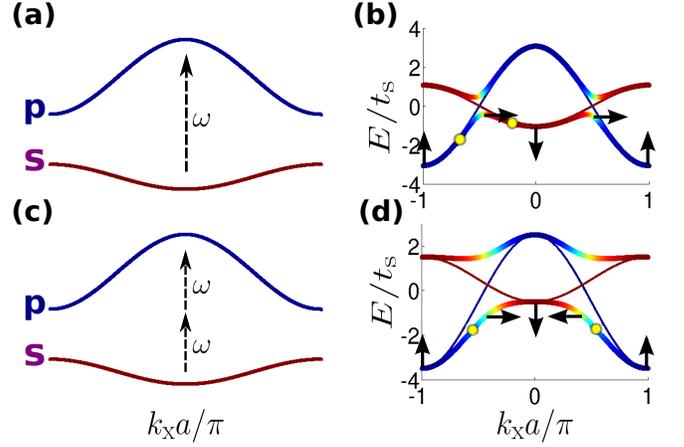}
\caption{\textbf{Single-particle band structure} \textbf{for }$t_{p}=3t_{s}$%
. (a) Illustration of one-photon coupling process of the two Bloch bands.
(b) Band structure of shaken lattice with finite one-photon coupling $\Omega
=0.3t_{s}$. (c) Illustration of the two-photon coupling process. (d) Band
structure of shaken lattice with finite two-photon coupling $\protect\alpha %
=1.0t_{s}$ and Zeeman field $h=0.3t_{s}$. To show how the coupling of two
Bloch bands opens an energy gap, we also plot the band dispersion without
coupling by the thin lines in (b,d). The colors in (b,d) represent the spin
compositions of each momentum state and the arrows represent the spin
rotations along one of the hybrid band. The yellow filled circles denote the
preferred Cooper pairings.}
\label{fig-band_structure}
\end{figure}

\textbf{Model Hamiltonian of the shaken lattice:} We first consider a
degenerate spinless Fermi gas trapped in a three dimensional (3D) optical
lattice. The system can be easily reduced to a quasi-2D or quasi-1D optical
lattice by raising the lattice potential depths along the $y$ and $z$
directions, allowing only small transverse tunneling. The shaking of the
lattice is along the $x$ direction and the shaking amplitude is ramped up
slowly to reach a constant \cite{Chin}, yielding a periodically modulating
lattice potential%
\begin{eqnarray}
V(x,y,t)&=&V_{x}\cos ^{2}\Big(k_{L}x+f\cos (\omega t)\Big)  \notag \\
&& +V_{y}\cos ^{2}\Big(k_{L}y\Big)+V_{z}\cos ^{2}\Big(k_{L}z\Big)~,
\end{eqnarray}%
where $V_{i}$ ($i=x,y,z$) are the lattice depths, $k_{L}=\pi /a$, $a$ is the
lattice spacing that is set as the length unit. $f$ and $\omega $ are the
shaking amplitude and frequency, respectively. The energy dispersions of the
static Bloch bands can be shifted by $n\hbar \omega $ ($n$ is an arbitrary
integer) due to the shaking, forming the new Floquet bands. The shaking also
couples two close Floquet or static bands, leading to gaps in the energy
spectrum, as illustrated in Fig. \ref{fig-band_structure}. We focus on the
lowest $s$- and $p$-bands and ignore all other higher bands for simplicity.
Because the shaking is along the $x$ direction, only the $p_{x} $-band can
be coupled with the \textit{s}-band and atoms stay at the $s$-band along the
other two directions. When $\omega $ is tuned close to the band gap ($\Gamma
$) of the static lattices, i.e, $\hbar \omega \sim \Gamma $, the $s$-orbital
state can absorb an energy of $\hbar \omega $ and couple with the $p_{x}$%
-orbital state, similar as the Rabi oscillation between two spin states.
Such \textquotedblleft one-photon process" coupling strength is denoted by $%
\Omega $, which can be approximated as a constant~\cite{QiZhou}. If we tune $%
\hbar \omega \sim \Gamma /2$, the $s$-orbital band is shifted upward by two
photon energy $2\hbar \omega $ to couple with the $p_{x}$-band. In this
\textquotedblleft two-photon process", by properly tuning the frequency and
amplitude of the shaking, the dominate coupling is between the nearest
neighboring sites between $s$- and $p_{x}$-orbital states that simulates a
SOC~\cite{QiZhou}.

In the basis of $(\psi _{s}(\bm{k}),\psi _{p_{x}}(\bm{k}))^{T}$, the
single-particle Hamiltonian in momentum space reads
\begin{equation}
\mathcal{H}_{0}=%
\begin{pmatrix}
\epsilon _{s}(\bm{k})+h & \Omega +\alpha \sin (k_{x}a) \\
\Omega +\alpha \sin (k_{x}a) & \epsilon _{p}(\bm{k})-h%
\end{pmatrix}
\label{Ham}
\end{equation}%
under the tight-binding approximation. Here $\epsilon _{s}(\bm{k}%
)=-t_{s}\cos (k_{x}a)-t_{s}^{\bot }\left[ \cos (k_{y}a)+\cos (k_{z}a)\right]
-\mu $ and $\epsilon _{p}(\bm{k})=t_{p}\cos (k_{x}a)-t_{s}^{\bot }\left[
\cos (k_{y}a)+\cos (k_{z}a)\right] -\mu $, where $t_{s}$ and $t_{p}$ are the
nearest neighbor tunneling amplitudes for an atom in $s$-orbital and $p$%
-orbital states along the $x$ direction, and $t_{s}^{\bot }$ is the
tunneling amplitudes along the $y$ and $z$ directions. $\Omega $ is the
momentum-independent coupling strength of the two orbital states which
dominates in the one-photon process, $\alpha $ is the effective SOC strength
which dominates in the two-photon process. $\mu $ is the chemical potential,
$h$ is the off-resonance detuning determined by the difference of the
shaking frequency and the band gap. Fig. \ref{fig-band_structure}
illustrates the one-photon and two-photon processes and the corresponding
band structures. With a finite coupling $\Omega $ or $\alpha \sin (k_{x}a)$,
two bands are hybridized around the crossing points, and thus yield an
energy gap. Note that for a 1D system with $t_{s}^{\bot }=0$ and $t_{p}=t_{s}
$, this non-interacting Hamiltonian breaks the time reversal symmetry but
preserves a chiral symmetry $\sigma _{y}\mathcal{H}_{0}(k_{x})\sigma _{y}=-%
\mathcal{H}_{0}(k_{x})$, which realizes an AIII class TI characterized by a $%
\mathbb{Z}$ topological invariant~\cite{Class1,Class2}. In the topological
phase, there are in-gap topological states on the boundaries of the system
\cite{AIII-TI-1, AIII-TI-2, BDI-Liu}. When $t_p\neq t_s$, there will be an
additional kinetic energy term, which does not change the phase transition
and topological properties of the system~\cite{Konig,HuiZhai}.

We consider on-site attractive interaction between atoms on different
orbital bands (see Discussion section and Supplementary Information for more
discussions). Such density-density interaction is not modified by the
shaking. In the momentum space, the interaction Hamiltonian can be written
as $\mathcal{H}_{I}=-~U\sum \psi _{s}^{\dag }(\bm{k}_{1})\psi _{p_{x}}^{\dag
}(\bm{k}_{2})\psi _{p_{x}}(\bm{k}_{3})\psi _{s}(\bm{k}_{4})$, where $\bm{k}%
_{1}+\bm{k}_{2}=\bm{k}_{3}+\bm{k}_{4}$ due to the momentum conservation in
the two-body scattering processes, $U>0$ is the interaction strength.

As the first step approach for a qualitative understanding of the
interacting Fermi gas in a shaken optical lattice, we consider the
mean-field approximation and assume a single-plane-wave FF-type order
parameter, \textit{i.e.}, $\Delta =U\langle \psi _{p_{x}}(\frac{\bm{Q}}{2}-%
\bm{k})\psi _{s}(\frac{\bm{Q}}{2}+\bm{k})\rangle $, where $\mathbf{Q}=\left(
Q,0,0\right) $ is the FF vector along the $x$ direction. $\bm{Q}=0$
corresponds to a conventional BCS superfluid. The dynamics of the system is
governed by the Bogliubov-de Gennes (BdG) Hamiltonian,
\begin{equation}
\mathcal{H}_{\mathrm{BdG}}(\bm{k})=\left(
\begin{array}{cc}
\mathcal{H}_{0}(\frac{\bm{Q}}{2}+\bm{k}) & \Delta \\
\Delta ^{\ast } & -\sigma _{y}\mathcal{H}_{0}^{\ast }(\frac{\bm{Q}}{2}-\bm{k}%
)\sigma _{y}%
\end{array}%
\right) ~,  \label{H_BdG}
\end{equation}%
in the Nambu-Gorkov spinor basis $[\psi _{s}(\frac{\bm{Q}}{2}+\bm{k}),\psi
_{p_{x}}(\frac{\bm{Q}}{2}+\bm{k}),\psi _{p_{x}}^{\dag }(\frac{\bm{Q}}{2}-%
\bm{k}),-\psi _{s}^{\dag }(\frac{\bm{Q}}{2}-\bm{k})]^{T}$. The gap and
momentum equations are solved by minimizing the thermodynamic potential to
obtain $\Delta $ and $Q$ (see Methods), through which we determine different
phases. When $\Delta \neq 0$ and $Q\neq 0$, the system is in a FFLO phase.
When $\Delta \neq 0$, $Q=0$, the system is in a BCS phase. Otherwise, the
system is a normal gas or a band insulator.

\begin{figure}[tbp]
\centering\includegraphics[width=0.48\textwidth]{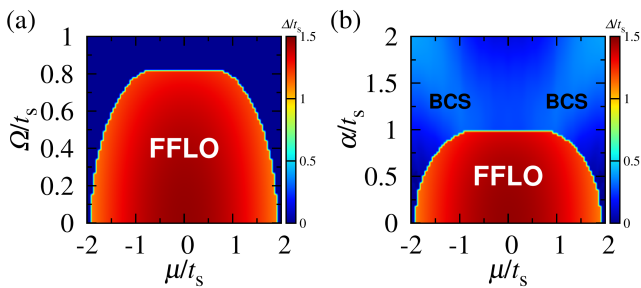}
\caption{\textbf{Phase diagram of a 3D shaken optical lattice with (a)
one-photon coupling $\Omega $ or (b) two-photon coupling $\protect\alpha $}.
The color describes the magnitude of the order parameter $\Delta $ in unit
of $t_{s}$. FFLO denotes the superfluid with a finite center of mass
momentum along the shaking direction $Q_{x}=\pm \protect\pi $, BCS stands
for the BCS superfluids with $\mathbf{Q=0}$. Other parameters $U=4.0t_{s}$, $%
h=0.0$, $t_{\text{p}}=3t_{\text{s}}$, $t_{s}^{\bot }=t_{s}$.}
\label{fig-2}
\end{figure}

\textbf{Phase diagrams in 3D lattices:} In Fig. \ref{fig-2} we plot the
phase diagrams for resonant one-photon (Fig. \ref{fig-2}a) and two-photon
processes (Fig. \ref{fig-2}b) with $t_{p}=3t_{s}$, $t_{s}^{\bot }=t_{s}$.
The phase diagram is similar for quasi-2D and quasi-1D systems with $%
t_{s}^{\bot }\rightarrow 0$. The system favors FFLO states in a large
parameter regime with a finite momentum along the shaking direction $%
Q_{x}=\pm \pi $. Here the FFLO pairing originates from the intrinsic band
dispersion inversion between $s$- and $p_{x}$-bands, which suppresses the
conventional BCS pairing. This can be further understood through a
coordinate transformation (see Methods) $k_{x}^{\prime }\rightarrow
k_{x}^{\prime }\pm \pi $ for the $p_{x}$-band dispersion to remove the band
inversion. In the system without the band inversion, we expect a
conventional BCS superfluids between $k_{x}$ and $-k_{x}$, leading to $%
k_{x}^{\prime }\pm \pi =-k_{x}$. Therefore the preferred pairings in the
shaken lattice are between two spins with momenta of $k_{x}$ and $%
k_{x}^{\prime }=\pm \pi -k_{x}$, and the FFLO momentum is $Q_{x}=\pm \pi $.
The analysis still applies in the presence of SOC and Zeeman fields.

For a large $\Omega $, the gap between two hybrid bands is very large, thus
a band insulator phase appears near the half filling. Each of the hybrid
band polarizes to one spin state for a large $\Omega $, leading to vanishing
Cooper pairing. In the presence of small SOC $\alpha \sin (k_{x}a)$, the
system still favors the FFLO superfluid. However, the SOC $\alpha \sin
(k_{x}a)$ leads to different wavefunctions at $k_{x}$ and $-k_{x}$, making
it possible to have the BCS pairing, as observed in Fig. \ref{fig-2}b for a
large SOC.

\begin{figure}[tbp]
\centering\includegraphics[width=0.48\textwidth]{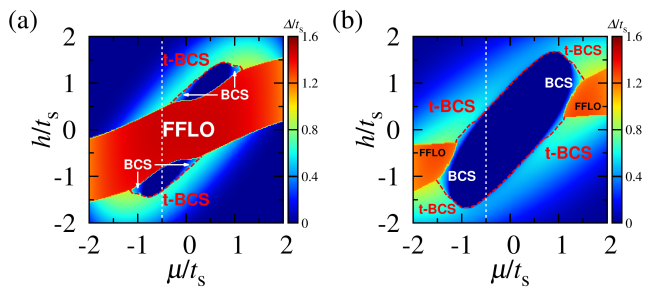}
\caption{\textbf{Effect of the Zeeman field on the phase diagram of the 1D
shaken optical lattice with a two-photon coupling}. (a) $\protect\alpha %
=0.7t_{s}$; (b) $\protect\alpha =1.4t_{s}$. t-BCS represents the topological
BCS superfluids. The red dashed lines are the boundary between BCS/insulator
and t-BCS superfluids. $U=4.0t_{s}$, $t_{\text{p}}=3t_{\text{s}}$, $\Omega
=0.0$. }
\label{fig-phase_diagram}
\end{figure}

\textbf{Topological phase and MFs in 1D lattices}: Hereafter we focus on
possible topological phases induced by the SOC. It is well known that there
is no topological phases in a 3D system with such 1D SOC. To reach the
topological phase that may support topological excitations such as MFs, we
need to consider a quasi-1D system with small\ or vanishing transverse
tunneling $t_{s}^{\bot }$. We first present the results for $t_{s}^{\bot }=0$
for simplicity, and will discuss how a small $t_{s}^{\bot }$ modifies the
phase diagram and the topological phases later. In Fig. \ref%
{fig-phase_diagram}, we plot the phase diagrams in the presence of a Zeeman
field $h$ for two different SOC strengths, where a new phase, topological
BCS (t-BCS) superfluids that host MFs, emerges in a large parameter regime.
The transition from BCS to t-BCS is characterized by the bulk quasiparticle
excitation spectrum closing and reopening at $k_{x}=0$ (and $k_{x}=\pm \pi $%
) and can be understood from the symmetry of the BdG Hamiltonian. The BdG
Hamiltonian (\ref{H_BdG}) satisfies the particle-hole symmetry $\Xi \mathcal{%
H}_{\mathrm{BdG}}(k_{x})\Xi ^{-1}=-\mathcal{H}_{\mathrm{BdG}}(-k_{x})$,
where $\Xi =\Lambda \mathcal{K}$, $\Lambda =\sigma _{x}\tau _{z}$ and $%
\mathcal{K}$ is the complex conjugate operator. For a BCS superfluid ($%
Q_{x}=0$), it also respects a time-reversal-like symmetry $\mathcal{T}%
\mathcal{H}_{\mathrm{BdG}}(k_{x})\mathcal{T}^{-1}=\mathcal{H}_{\mathrm{BdG}%
}(-k_{x})$ if $t_{p}=t_{s}$, where $\mathcal{T}=\sigma _{z}\tau _{0}\mathcal{%
K}$. This topological BCS superfluid belongs to the BDI symmetry class
characterized by a $\mathbb{Z}$ invariant and MFs can be found at the
boundary of the superfluids~\cite{Class1,Class2,LiuXJ}. For $t_{p}\neq t_{s}$%
, it belongs to the more generalized D symmetry class characterized by a $%
\mathbb{Z}_{2}$ invariant. The topological BCS phase region can be
determined by the Pfaffian sign of the skew matrix $\Gamma (k_{x})=H_{\text{%
BdG}}(k_{x})\Lambda $ yielding
\begin{equation}
\mathrm{sign}[\mathrm{Pf}(\Gamma (0))\times \mathrm{Pf}(\Gamma (\pi ))]=-1~,
\label{pfaffian}
\end{equation}%
which has an explicit form
\begin{equation}
[\left(t_{+}-h\right)^{2}-\Delta ^{2}-\left( \mu +t_{-}\right) ^{2}] [\left(
t_{+}+h\right) ^{2}-\Delta ^{2}-\left( \mu -t_{-}\right) ^{2}]<0\noindent
\label{PFexpression}
\end{equation}%
where $t_{+}=(t_s+t_p)/2$, $t_{-}=(t_s-t_p)/2$.

Because $t_{\text{p}}\neq t_{\text{s}}$, the phase diagrams are not
symmetric about $h=0$ or $\mu =0$ as shown in Fig. \ref{fig-phase_diagram}.
However, the system is symmetric with the transformation $h\rightarrow -h$
and $\mu \rightarrow -\mu $. This is in stark contrast to conventional
systems (two pseudo-spins both in $s$-bands), where the phase diagrams are
symmetric with respect to either $h=0$ or $\mu =0$ and the Zeeman field must
be larger than a critical value for the appearance of t-BCS phase.
Furthermore, when the inversion symmetry of the hybrid band structure is
broken, it is also possible to realize the topological FFLO superfluids in
the shaken optical lattice~\cite{topoFF1, topoFF2,topoFF3,topoFF4}.

From Fig.\ref{fig-phase_diagram}, we see the FFLO superfluid dominates for
small SOC and the region of t-BCS superfluids does not change much when the
strength of SOC is increased. There exists an insulator block with $\Delta
=0 $ near $\mu =0$ surrounded by the superfluid phase in Fig. \ref%
{fig-phase_diagram}b. Note that without the interaction, the single particle
Hamiltonian $\mathcal{H}_{0}$ (with $t_{s}^{\bot }=0$) already possesses
some topological properties in the presence of the SOC. In this insulator
phase, a pair of sub-gap states may appear on the system boundaries~\cite%
{HuiZhai}. With the interaction tuned on in Fig.\ref{fig-phase_diagram}%
(a,b), the system evolves into a topological superfluids with finite BCS
order parameters. The original edge states of the topological insulator are
now replaced with the zero energy Majorana boundary states \cite{BDI-Liu}
when the chemical potential is in the band gap, similar as previous TI based
MF scheme using edge states. More interestingly, when the chemical potential
cuts only one of the conduction or valance bands, we find the coexistence of
the edge states from the topological insulator and the zero energy Majorana
edge states from the topological superconductor.

For MFs, we can consider 1D atom tubes generated by optical lattices in a 3D
system with weak tunnelings along the transverse directions. The weak
transverse tunneling can strongly suppress the quantum fluctuations along
the 1D tubes, similar as that in high temperature cuprate superconductors.
Consider a periodic boundary condition along the transverse directions, the
weak tunneling simply shifts the chemical potential of the 1D gas at most by
4$t_{\bot }$ in Eq. (\ref{PFexpression}). As long as the shifted chemical
potential still stays inside the topological BCS region, we expect the MFs
exist along the tube edges. Similar issue for MFs has been widely discussed
in spin-orbit coupled quantum wires (nanowire or cold atom tube arrays) and
our calculations show that the same conclusion still holds for the shaken
optical lattices.

\begin{figure}[tbp]
\centering\includegraphics[width=0.48\textwidth]{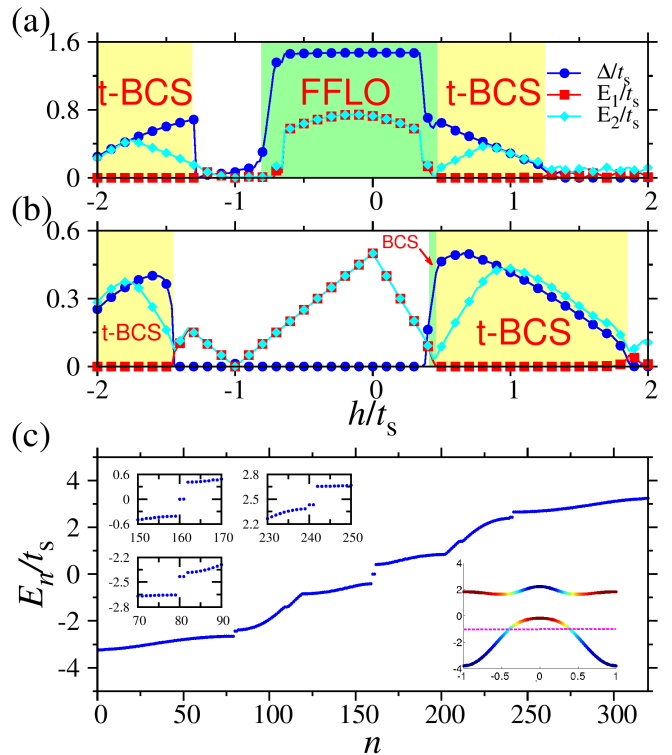}
\caption{\textbf{Majorana fermions in a shaken optical lattice}. (a, b) The
order parameter $\Delta $ (Blue circles) and the lowest two quasi-particle
excitation energies $E_{1}$ (red squares), $E_{2}$ (cyan diamonds) as a
function of Zeeman field $h$ by self-consistently solving the real space BdG
equation with open boundary condition for $\protect\alpha =0.7t_{\text{s}}$
(a) and $\protect\alpha =1.4t_{\text{s}}$ (b). We have labeled the phase
region for FFLO superfluids and the topological BCS superfluids with colors.
Other unlabeled regions correspond to either normal gas or insulator phase.
In both figures $\protect\mu =-0.5t_{s}$, $U=4.0t_{s}$, $\Omega =0.0$, $t_{%
\text{p}}=3t_{\text{s}}$. (c) Plot of the BdG quasi-particle excitation
energies for $\protect\alpha =1.4t_{s}$, $\protect\mu =-0.7t_{s}$, $%
h=0.8t_{s}$. There are three pairs of sub-gap states all localized on the
boundaries.}
\label{fig-majorana}
\end{figure}

\textbf{Quantum phase transition and Majorana fermions in real space:} The
above momentum space analysis is further confirmed by self-consistently
solving the BdG Hamiltonian in the real space. In Fig. \ref{fig-majorana}%
(a,b) we plot the average value of the order parameter $\Delta $ and the
lowest two quasi-particle excitation energies $E_{1}$ and $E_{2}$ as a
function of the Zeeman field $h$ for (a) $\alpha =0.7t_{s}$ and for (b) $%
\alpha =1.4t_{s}$. In consistent with the white dashed lines in Fig. \ref%
{fig-phase_diagram}, there is a phase transition from FFLO superfluids to
topological BCS superfluids at $h\approx 0.48t_{s}$ for small SOC in (a) and
from insulator phase to trivial BEC superfluids and then to a topological
BCS superfluids at $h\approx 0.45t_{s}$ for large SOC in (b). The momentum
of FFLO Cooper pairs in real space is either FF or LO types, both have the
same ground state energies as that in momentum space. In the topological
phase, the zero energy Majorana fermions are protected by a finite mini-gap $%
E_{2}$. In Fig. \ref{fig-majorana}(c) we plot of the BdG quasi-particle
excitation energies for $\mu =-0.7t_{s}$, $h=0.8t_{s}$, $\alpha =1.4t_{s}$.
The inset shows the quasi-particle excitation spectrum with zero energy
degenerate MF states and the single particle band structure where the
chemical potential cuts a single conduction band. The zero energy Majorana
fermion modes are localized on the system boundaries. Interestingly for $\mu
\neq 0$ and when the chemical potential cuts either of the two bands, we may
find another one pairs of sub-gap edge states with finite energies which are
induced by the topological insulator. The coexistence of the reminiscent
edge states from topological insulator and the MFs edge states from
topological superconductor may lead to many interesting transport properties
in this system.

{\LARGE \textbf{Discussion}}

For a spinless Fermi gas, the $s$-wave scattering interactions between same
orbital states are usually prohibited by the Pauli exclusion principle, and
the interactions between different orbital states are usually very small.
However, the strong on-site attractive interactions can still be engineered
using different methods (see more detailed discussion in Supplementary
Information):

a) Bosons-mediated interactions in Bose-Fermi mixtures \cite{Pethick,mixture}%
, where the induced interaction between fermions by bosons can be written as
screened Coulomb interaction $U_{ind}\left( r\right) =-\frac{%
m_{B}n_{B}U_{BF}^{2}}{\pi \hbar ^{2}}\frac{\exp \left( -\sqrt{2}r/\xi
\right) }{r}$. Here $m_{B}$ and $n_{B}$ are the mass and density of the
bosons, $\xi =\sqrt{\hbar ^{2}/2m_{B}n_{B}U_{BB}}$ is the healing length for
the bosons, and $U_{BF}$ and $U_{BB}$ are the Bose-Fermi and Bose-Bose
scattering interaction strengths that can be tuned using Feshbach resonance;

b) Dipole-dipole interaction between dipolar atoms~\cite{CJWu}: when the
dipoles are aligned along the 1D direction (i.e., head to head), the dipole
interaction between atoms ($\sim 1/r^{3}$) is attractive. Both the screened
Coulomb interaction in Bose-Fermi mixture and the long rang dipole
interaction can lead to attractive on-site interaction between atoms with
different orbital states at the same lattice site. For the dipolar atoms,
the on-site interaction can be tuned by the lattice confinement along the
transverse directions.

c) Through the interaction between the atom in the $p$-orbital with another $%
s$-orbital atom with different hyperfine state or different specie, where
the interaction can also be tuned by Feshbach resonance. In this case, the
optical lattice should be spin-dependent or species dependent so that the $s$%
-orbital of the atom does not couple to the $p$-orbital for the same specie.
The alkali-earth or heavy alkali atoms can be used to avoid heating from
spontaneous emission. For different species (i.e., Fermi-Fermi mixtures),
far detuning lasers can be used without any significant heating. In this
system, the $s$-orbital of the original atoms are assumed to be fully
filled. Therefore even they interact with the additional atoms, they do not
generate superfluid order. Our calculations show that FFLO states can also
be generated in a certain parameter region.

\textbf{Acknowledgement} We thank Yong Xu, Chris Hamner and Peter Engels for
helpful discussions. C. Qu and C. Zhang are supported by ARO
(W911NF-12-1-0334) and AFOSR (FA9550-11-1-0313). Z.Z. and X.Z. are supported
by National Natural Science Foundation of China (Grant No. 11074244 and
Grant No. 11274295), and National 973 Fundamental Research Program
(2011cba00200).

{\LARGE \textbf{Methods}}

\textbf{Mean-field model in the momentum space.} Consider the order
parameter $\Delta (\bm{r})=U\langle \psi _{p}\psi _{s}\rangle =\Delta e^{i%
\bm{Q}\cdot \bm{r}}$, where $\bm{Q}$ is the pairing momentum, the
thermodynamical potential is given by
\begin{eqnarray}
\Omega &=&\dfrac{1}{2}\sum\limits_{\bm{k}}\Big(\epsilon _{s}(\bm{Q}/2-\bm{k}%
)+\epsilon _{p}(\bm{Q}/2-\bm{k})\Big)  \notag \\
&&+~\sum\limits_{\bm{k}\lambda }\Theta \Big(-E_{\lambda }(\bm{k})\Big)%
E_{\lambda }(\bm{k})+\dfrac{|\Delta |^{2}}{U}~,
\end{eqnarray}%
where $\Theta (x)$ is the Heaviside function representing the Fermi
distribution at zero temperature. $E_{\lambda }(\bm{k})$($\lambda =1,\cdots
,4$) are the four eigenvalues of the BdG Hamiltonian $\mathcal{H}_{\mathrm{%
BdG}}(\bm{k})$. The order parameter $\Delta $ and momentum $\bm{Q}$ are
hence given by self-consistently solving the saddle equations of the
thermodynamical potential $\Omega $:
\begin{equation}
\dfrac{\partial \Omega }{\partial \Delta }=0~,\quad \dfrac{\partial \Omega }{%
\partial \bm{Q}}=0~.
\end{equation}

\textbf{Mean-field model in the real space.} In the real space the
tight-binding Hamiltonian is written as
\begin{equation}
H_{\mathrm{TB}}=H_{0}+H_{\mathrm{Z}}+H_{\mathrm{\alpha }}+V_{\mathrm{int}}~,
\label{eq-TB}
\end{equation}%
where
\begin{equation}
H_{0}=\sum_{\langle i,j\rangle }\Big(-t_{s}c_{i}^{\dagger
}c_{j}+t_{p}c_{i}^{\dagger }c_{j}\Big)-\mu \sum_{i,\sigma =s,p}n_{i\sigma },
\end{equation}%
\begin{equation}
H_{\mathrm{Z}}=-h_{z}\sum_{i}(n_{is}-n_{ip}),
\end{equation}%
\begin{equation}
H_{\mathrm{\alpha }}=\frac{\alpha }{2}\sum_{i}(c_{i-1p}^{\dagger
}c_{is}-c_{i+1p}^{\dagger }c_{is}+\mathrm{H.C.}),
\end{equation}%
\begin{equation}
V_{\mathrm{int}}=-U\sum_{i}n_{is}n_{ip},
\end{equation}%
The mean field order parameters
\begin{equation}
\Delta _{i}=U\langle c_{ip}c_{is}\rangle .
\end{equation}%
We have defined $c_{i}$ and $c_{i}^{\dagger }$ as the particle annihilation
and creation operator on site $i$ and the particle number operator $%
n_{i\sigma =s,p}=c_{i\sigma }^{\dagger }c_{i\sigma }$.

\textbf{Mechanism of FFLO pairing.} To demonstrate why the system favors
FFLO pairing with a finite momentum $Q=\pm \pi /a$ along the SOC direction,
we start with the most simplest case, i.e., with vanishing $\alpha $, $h$
and $\mu $ and consider only the 1D system. If we choose the Nambu-Gorkov
spinor $\Psi =\Big(\psi _{\text{s}}(k_{x}),\psi _{p}(k_{x}^{\prime }),\psi
_{p}^{\dag }(k_{x}^{\prime }),\psi _{\text{s}}^{\dag }(k_{x})\Big)^{T}$, and
introduce a general pairing order parameter $\Delta =U\langle \psi _{p}\psi
_{\text{s}}\rangle $ the BdG Hamiltonian $\mathcal{H}_{\mathrm{BdG}}(k_{x})$
is rewritten as
\begin{equation}
\mathcal{H}_{\mathrm{BdG}}(k_{x},k_{x}^{\prime })=\left(
\begin{array}{cc}
T(k_{x},k_{x}^{\prime }) & \Delta \\
\Delta ^{\ast } & -\sigma _{y}T(k_{x},k_{x}^{\prime })\sigma _{y}%
\end{array}%
\right) ~,
\end{equation}%
where $T(k_{x},k_{x}^{\prime })=\mathrm{diag}[-t_{\text{s}}\cos
(k_{x}a),t_{p}\cos (k_{x}^{\prime }a)]$. When transferred to a new spinor
basis $\Psi ^{\prime }=\Big(\psi _{\text{s}}(k_{x}),\psi _{p}(-\frac{\pi }{a}%
+k_{x}^{\prime }),\psi _{p}^{\dag }(-\frac{\pi }{a}+k_{x}^{\prime }),\psi _{%
\text{s}}^{\dag }(k_{x})\Big)^{T}$ through a unitary transformation $\Psi
=U\Psi ^{\prime }$ with $U=\mathrm{diag}(1,e^{i\pi x/a},e^{i\pi x/a},1)$, we
get
\begin{eqnarray}
&&\mathcal{H}_{\mathrm{BdG}}^{\prime }(k_{x},k_{x}^{\prime \dag }\mathcal{H}%
_{\mathrm{BdG}}(k_{x},k_{x}^{\prime })U  \notag \\
&=&\left(
\begin{array}{cc}
T(k_{x},k_{x}^{\prime }-\frac{\pi }{a}) & \Delta \\
\Delta ^{\ast } & -\sigma _{y}T(k_{x},k_{x}^{\prime }-\frac{\pi }{a})\sigma
_{y}%
\end{array}%
\right) ~.
\end{eqnarray}%
Notice that $T(k_{x},k_{x}^{\prime }-\frac{\pi }{a})=\mathrm{diag}[-t_{\text{%
s}}\cos (k_{x}a),-t_{p}\cos (k_{x}^{\prime }a)]$ correspond to the
conventional bands which are known to favor a BCS pairing in this new basis.
It leads to $k_{x}=-(\pm \pi /a+k_{x}^{\prime })$ and hence $%
k_{x}+k_{x}^{\prime }=\pm \pi /a$. As a result the pairing momentum should
be fixed to $\pm \pi /a$. \newline

{\LARGE Supplementary Information}

\begin{figure}[tbp]
\centering\includegraphics[width=0.48\textwidth]{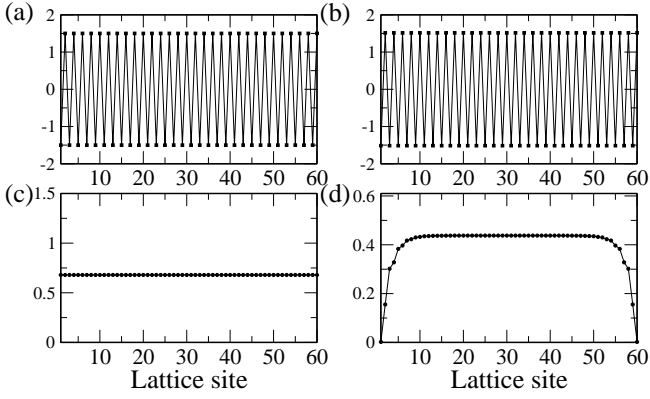}
\caption{\textbf{Self-consistently obtained order parameter in real space}.
(a) $\Omega =0.3t_{\text{s}}$, $\protect\mu =0.5t_{\text{s}}$, the system is
in FFLO phase; (b) $\protect\alpha =0.3t_{\text{s}}$, $\protect\mu =0.5t_{%
\text{s}}$, the system is in FFLO phase; (c) $\protect\alpha =1.2t_{\text{s}%
} $, $\protect\mu =1.2t_{\text{s}}$, the system is in trivial BCS phase; (d)
$\protect\alpha =1.2t_{\text{s}}$, $\protect\mu =1.2t_{\text{s}}$, $h=-0.5t_{%
\text{s}}$, the system is in topological BCS phase where we adopt open
boundary condition for the appearance of MFs. Other parameters: $U=4.0t_{%
\text{s}}$, $t_{\text{p}}=3t_{\text{s}}$. }
\label{fig-delta}
\end{figure}

\begin{figure}[tbp]
\centering\includegraphics[width=0.48\textwidth]{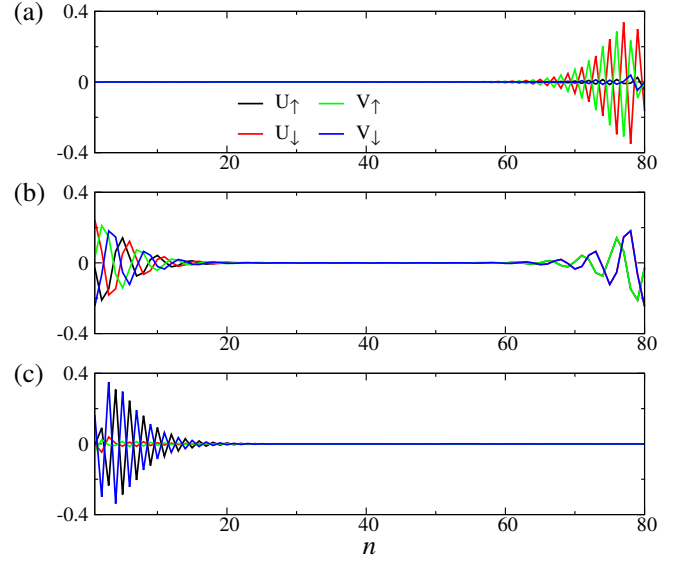}
\caption{\textbf{Wave functions of three pairs of edge states in real space}%
. (a,b,c) are the wave functions of the three pairs of sub-gap edge states
as shown in Fig. \protect\ref{fig-majorana} for energy level $E_{n}$ where $%
n=80,160,240$. (a,c) are the wave functions closely related to the
topological insulator when there are no interactions; (b) are the wave
functions of zero-energy Majorana fermions. Other parameters are $\protect%
\alpha =1.4t_{\text{s}}$, $\protect\mu =-1.0t_{\text{s}}$, $h=0.8t_{\text{s}%
} $, $U=4.0t_{\text{s}}$.}
\label{fig-edge}
\end{figure}

\textbf{Generation of on-site interactions.} For a three dimensional optical
lattice, each well can be expanded around its center as a harmonic
oscillator potential
\begin{equation}
V(\bm{r})=\dfrac{1}{2}m(\omega _{x}^{2}x^{2}+\omega _{y}^{2}y^{2}+\omega
_{z}^{2}z^{2})~,
\end{equation}%
where $\bm{r}=(x,y,z)$. We focus on the on-site interaction of the lowest
two orbital states, for instance $s$- and $p_{x}$-orbital states, and assume
an $s$-orbital state along both $y$- and $z$-axes. The wave functions for $s$%
- and $p_{x}$-orbital states in this harmonic potential are given by
\begin{eqnarray}
\psi _{s}(\bm{r}) &=&\Big(\dfrac{m}{\pi \hbar }\Big)^{\frac{3}{4}}\omega
_{x}^{\frac{1}{4}}\omega _{y}^{\frac{1}{4}}\omega _{z}^{\frac{1}{4}}  \notag
\\
&&\times \exp \Big[-\dfrac{m}{2\hbar }(\omega _{x}x^{2}+\omega
_{y}y^{2}+\omega _{z}z^{2})\Big]~,  \notag \\
\psi _{p}(\bm{r}) &=&\sqrt{2\pi }\Big(\dfrac{m}{\pi \hbar }\Big)^{\frac{5}{4}%
}\omega _{x}^{\frac{3}{4}}\omega _{y}^{\frac{1}{4}}\omega _{z}^{\frac{1}{4}}
\notag \\
&&\times x\exp \Big[-\dfrac{m}{2\hbar }(\omega _{x}x^{2}+\omega
_{y}y^{2}+\omega _{z}z^{2})\Big]~.  \notag \\
&&
\end{eqnarray}%
In the Hartree-Fock approximation, the on-site interaction can be evaluated
as,
\begin{eqnarray}
-U &=&\int d^{3}r_{1}d^{3}r_{2}~V(\bm{r}_{1}-\bm{r}_{2})\Big[|\psi _{s}(%
\bm{r}_{1})\psi _{p}(\bm{r}_{2})|^{2}  \notag \\
&&-\psi _{s}^{\ast }(\bm{r}_{1})\psi _{p}^{\ast }(\bm{r}_{2})\psi _{s}(\bm{r}%
_{2})\psi _{p}(\bm{r}_{1})\Big]  \notag \\
&=&\dfrac{2m^{4}}{\pi ^{3}\hbar ^{4}}\omega _{x}^{2}\omega _{y}\omega
_{z}\int d^{3}Rd^{3}r~V(\bm{r})r_{x}(r_{x}/2-R_{x})  \notag \\
&&\times \exp \Big[-\dfrac{m}{\hbar }\sum\limits_{i=x,y,z}\omega
_{i}(R_{i}^{2}+r_{i}^{2}/4)\Big]~,  \label{onsite_interaction}
\end{eqnarray}%
where we have introduced $\bm{r}=\bm{r}_{1}-\bm{r}_{2}$ and $\bm{R}=(\bm{r}%
_{1}+\bm{r}_{2})/2$. For simplicity, we assume $\omega _{y}=\omega
_{z}=\omega $ in following calculations. For the fermionic atomic gas with a
dipole moment $\bm{M}$, the anisotropic interaction is expressed as
\begin{equation}
V(\bm{r})=\dfrac{\mu _{0}|\bm{M}|^{2}}{4\pi }\dfrac{[1-3(\hat{M}\cdot \hat{r}%
)]^{2}}{r^{3}}~,
\end{equation}%
where $\hat{M}=\bm{M}/|\bm{M}|$ and $\hat{r}=\bm{r}/|\bm{r}|$. Substituting
it into the on-site interaction expression Eq. (\ref{onsite_interaction}),
we get
\begin{equation}
-U=2\Big(\dfrac{m\omega }{\pi \hbar }\Big)^{\frac{3}{2}}\dfrac{\mu _{0}|%
\bm{M}|^{2}}{4\pi }~\times F(\frac{\omega _{x}}{\omega })
\end{equation}%
where
\begin{equation}
F(\frac{\omega _{x}}{\omega })=\int_{0}^{2\pi }d\theta \int_{0}^{\pi }d\phi
\dfrac{(\frac{\omega _{x}}{\omega })^{\frac{3}{2}}\cos ^{2}\theta \sin
^{3}\phi ~[1-3(\hat{M}\cdot \hat{r})]^{2}}{(\frac{\omega _{x}}{\omega }\cos
^{2}\theta +\sin ^{2}\theta )\sin ^{2}\phi +\cos ^{2}\phi }~.  \notag
\end{equation}

In a real experiment, we can use fermionic atoms of $^{167}\mathrm{Er}$ with
$M=7\mu _{B}$. The optical lattice can be created by laser beams with the
wavelength $\lambda \approx 600nm$. The recoil energy of such an optical
lattice is $E_{R}=h^{2}/2m\lambda =157nK$. For a typical tunneling energy $%
t_{s}\approx 0.36E_{R}\approx 55nK$. The on-site interaction $U\approx 224nK$
when $\omega _{x}/\omega \approx 0.3$ hence $U\approx 4t_{\text{s}}$ adopted
in our paper can be obtained in real experiments.

\textbf{Order parameters from self-consistent calculation in real space.}
Our numerical calculations are done in both momentum space and real space,
which agrees very well with each other. In Fig. \ref{fig-delta}, we present
the order parameter profiles in various phases from real space calculations.

\textbf{Coexistence of edge states in the shaken optical lattice} Without
interaction, the system supports gap states which are localized on the
system boundaries; When the interaction is turned on and the chemical
potential cuts either the conduction or valance band, the system may support
multiple edge states as shown in Fig.~\ref{fig-edge}. The coexistence of the
reminiscent edge states from topological insulator and the MFs edge states
from topological superconductor may lead to many interesting transport
properties in this system.

\begin{figure}[tbp]
\centering\includegraphics[width=0.48\textwidth]{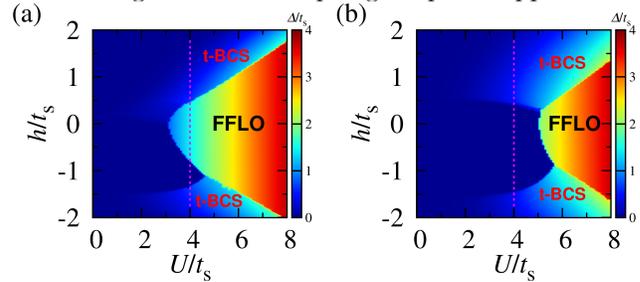}
\caption{\textbf{BCS-BEC crossover}. Color plots of the order parameter in
the $h-U$ plane for small SOC (a) $\protect\alpha =0.7t_{\text{s}}$ and
large SOC (b) $\protect\alpha =1.4t_{\text{s}}$. The chemical potential is
taken as $\protect\mu =-0.5t_{\text{s}}$, and $\Omega =0.0$, $t_{\text{p}%
}=3t_{\text{s}}$. }
\label{fig-crossover}
\end{figure}

\textbf{BCS-BEC crossover with two-photon coupling} The interaction strength
$U$ can be easily tuned in cold atom experiments which provides a way to
study the crossover from the BCS superfluids with weak attractive
interaction to the Bose-Einstein condensation (BEC) of strongly bounded
molecules. BCS-BEC crossover has been widely studied in ultracold Fermi
gases in various free space and optical lattice systems \cite%
{crossover1,crossover2}. In Fig. \ref{fig-crossover} we plot the phase
diagrams in $h$-$U$ plane for a two-photon coupling process with different
strengths of SOC (a) $\alpha =0.7t_{\text{s}}$ (b) $\alpha =1.4t_{\text{s}}$
for a certain value of chemical potential. In consistent with Fig. \ref%
{fig-phase_diagram} (a)(b) we see FFLO superfluid phase dominates for small
SOC and small Zeeman field, and BCS phase dominates for large Zeeman field
or large SOC. The topological phase appears when the strength of the Zeeman
field exceeds a critical value for a medium value of interaction strength $U$
and tends to disappear at the BEC side. Because SOC leads to an energy gap
near half filling, the band insulator occupies a larger parameter region
when SOC coupling strength is increased in the BCS side.

\begin{figure}[tbp]
\centering\includegraphics[width=0.48\textwidth]{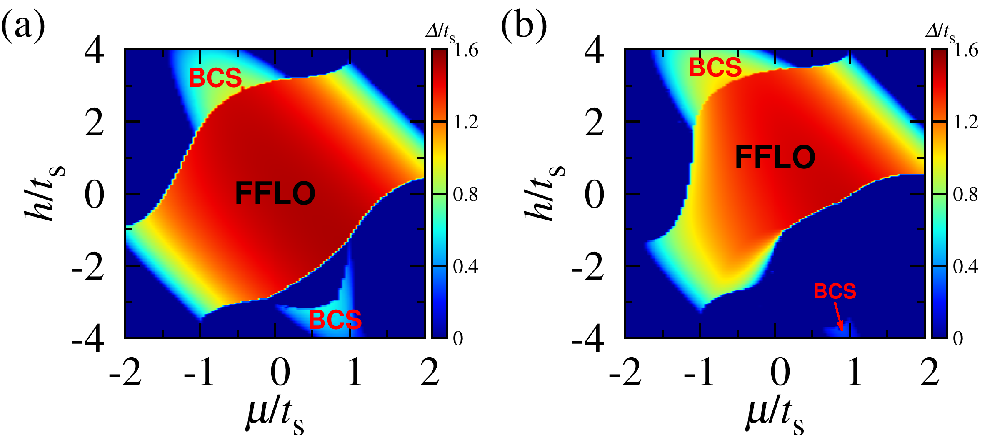}
\caption{\textbf{Effect of the Zeeman field on the phase diagram of the
shaken optical lattice with the interaction with additional specie of atoms
for a two-photon coupling}. (a) $\protect\alpha =0.7t_{s}$; (b) $\protect%
\alpha =1.4t_{s}$. We set $U=4.0t_{s}$, $t_{\text{p}}=3t_{\text{s}}$, $%
\Omega =0.0$. }
\label{tri-2}
\end{figure}

\begin{figure}[tbp]
\centering\includegraphics[width=0.48\textwidth]{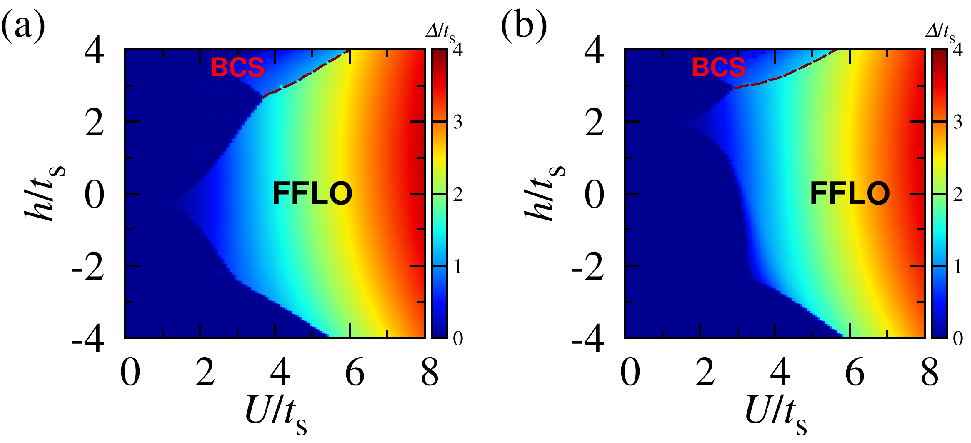}
\caption{\textbf{BCS-BEC crossover with the interaction with additional
specie of atoms}. Color plots of the order parameter in the $h-U$ plane for
smaller SOC (a) $\protect\alpha =0.7t_{\text{s}}$ and larger SOC (b) $%
\protect\alpha =1.4t_{\text{s}}$. The chemical potential is taken as $%
\protect\mu =-0.5t_{\text{s}}$, and $\Omega =0.0$, $t_{\text{p}}=3t_{\text{s}%
}$. }
\label{tri-3}
\end{figure}

\textbf{Spin $1/2$ Fermi gas in a shaken optical lattice.} We study the
phase diagrams of a spin $1/2$ (two internal states) Fermi gas in a shaken
optical lattice, where atoms in one internal state only occupy the $s$%
-orbital band, and atoms in the other internal state occupy both $s$- and $p$%
-orbital band. In the shaking, the $s$- and $p$-orbital bands of the same
species atoms are hybridized, and we only assume the interactions between
different species in $s$- and $p$-orbital states. In the basis of $(\psi
_{p\uparrow }(k_{1}),\psi _{s\uparrow }(k_{1}),\psi _{s\downarrow
}(k_{1}),\psi _{s\downarrow }^{\dag }(k_{2}),\psi _{s\uparrow }^{\dag
}(k_{2}),-\psi _{p\uparrow }^{\dag }(k_{2}))^{T}$ with $k_{1}=Q/2+k_{x}$ and
$k_{2}=Q/2-k_{x}$, the Hamiltonian in momentum space reads
\begin{widetext}%
\begin{equation}%
\mathcal{H}=%
\begin{pmatrix}
\varepsilon_{p\uparrow}(k_1)-h  & \alpha \sin (k_1a) & 0 & \Delta & 0 & 0 \\
\alpha \sin (k_1a) & \varepsilon_{s\uparrow}(k_1) & 0 & 0 & 0 & 0 \\
0 & 0 & \varepsilon_{s\downarrow}(k_1) & 0 & 0 & \Delta \\
\Delta & 0 & 0 & -\varepsilon_{s\downarrow}(k_2) & 0 & 0 \\
0 & 0 & 0 & 0 & -\varepsilon_{s\uparrow}(k_2) & \alpha \sin (k_2a) \\
0 & 0 & \Delta & 0 & \alpha \sin (k_2a) & -\varepsilon_{p\uparrow}(k_2)+h
\end{pmatrix}%
\end{equation}%
\end{widetext}where $\varepsilon _{p\uparrow }(k_{x})=t_{p}\cos (k_{x}a)-\mu
$, $\varepsilon _{s\uparrow }(k_{x})=-t_{s}\cos (k_{x}a)-\mu _{s\uparrow }$
and $\varepsilon _{s\downarrow }(k_{x})=-t_{s}\cos (k_{x}a)-\mu $. For spin $%
\uparrow $ in $s$-orbit states, we assume that it is full occupied with $\mu
_{\text{s}\uparrow }=t_{\text{s}}$. Hence the interaction between the two
different species in the same $s$-orbits states can be ignored. Fig. \ref%
{tri-1},\ref{tri-2},\ref{tri-3} demonstrate that similar phase diagrams are
obtained where we identify the FFLO and BCS superfluids.

\begin{figure}[tbp]
\centering\includegraphics[width=0.48\textwidth]{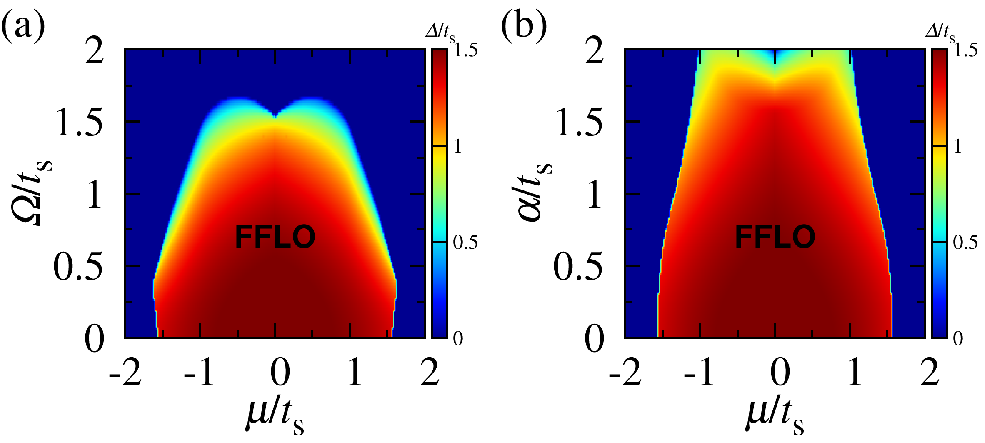}
\caption{\textbf{Phase diagram of the shaken optical lattice with the
interaction with additional species of atoms (a) one-photon coupling $\Omega
$ or (b) two-photon coupling $\protect\alpha $}. The color describes the
magnitude of the order parameter $\Delta $ in unit of $t_{\text{s}}$. Other
parameters $U=4.0t_{\text{s}}$, $h=0.0$, $t_{\text{p}}=3t_{\text{s}}$.}
\label{tri-1}
\end{figure}

\begin{figure}[tbp]
\centering\includegraphics[width=0.48\textwidth]{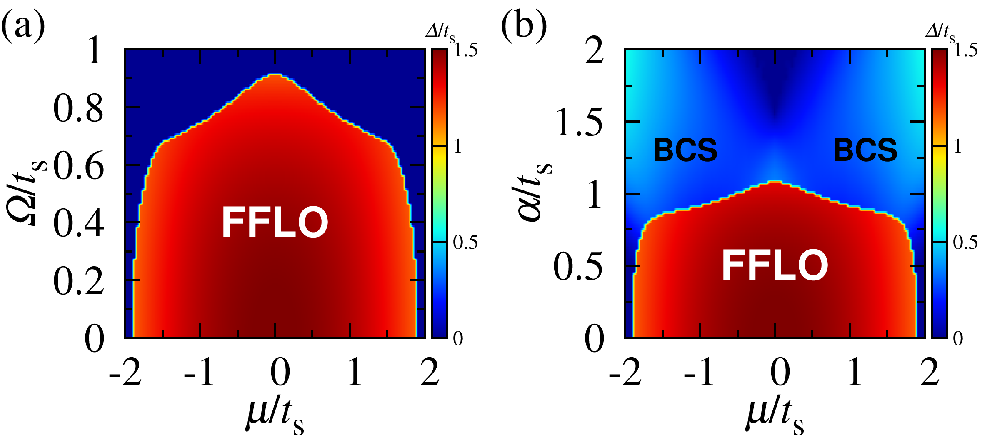}
\caption{\textbf{Phase diagram of the 2D shaken optical lattice with (a)
one-photon coupling $\Omega $ and (b) two-photon coupling $\protect\alpha $}%
. The color describes the magnitude of the order parameter $\Delta $ in unit
of $t_{\text{s}}$. Other parameters $U=4.0t_{\text{s}}$, $h=0.0$, $t_{\text{p%
}}=3t_{\text{s}}$, $t_{s}^{\perp }=t_{s}$.}
\label{2d-case}
\end{figure}

\textbf{FFLO superfluids in a 2D shaken optical lattice.} The FFLO
superfluids induced by the shaken optical lattice also exist in a 2D case.
For simplicity, we consider the one-photon coupling $\Omega $ between the
lowest two orbital states and assume the Cooper pairing can only have a
nonzero momentum along $x$ direction. In the basis of $(\psi _{p_{x}}(\bm{k}%
_{1}),\psi _{s}(\bm{k}_{1}),\psi _{s}^{\dag }(\bm{k}_{2}),-\psi
_{p_{x}}^{\dag }(\bm{k}_{2}))^{T}$ with $\bm{k}_{1}=(Q/2+k_{x},k_{y})$ and $%
\bm{k}_{2}=(Q/2-k_{x},-k_{y})$, the Hamiltonian reads
\begin{equation}
\mathcal{H}=%
\begin{pmatrix}
\varepsilon _{p_{x}}(\bm{k}_{1}) & \Omega & \Delta & 0 \\
\Omega & \varepsilon _{s}(\bm{k}_{1}) & 0 & \Delta \\
\Delta & 0 & -\varepsilon _{s}(\bm{k}_{2}) & \Omega \\
0 & \Delta & \Omega & -\varepsilon _{p_{x}}(\bm{k}_{2})%
\end{pmatrix}%
\end{equation}%
where $\varepsilon _{p_{x}}(\bm{k})=-t_{p}\cos (k_{x}a)+t_{s}^{\perp }\cos
(k_{y}a)-\mu $ and $\varepsilon _{s}(\bm{k})=t_{s}\cos (k_{x}a)+t_{s}^{\perp
}\cos (k_{y}a)-\mu $. The phase diagram is shown in Fig. \ref{2d-case}.

\begin{thebibliography}{99}
\bibitem{Bloch-NP} Bloch, I., Dalibard, J. \& Nascimbene, S. Quantum
simulations with ultracold quantum gases. \textit{Nature Physics} \textbf{8,}
267-276 (2012).

\bibitem{Jaksch-1998} Jaksch, D., Bruder, C., Cirac, J. I., Gardiner, C. W.
\& Zoller, P. Cold bosonic atoms in optical lattices. \textit{Phys. Rev.
Lett.} \textbf{81}, 3108-3111 (1998).

\bibitem{Bloch-RMP} Bloch, I., Dalibard, J. \& Zwerger, W. Many-body physics
with ultracold gases. \textit{Rev. Mod. Phys.} \textbf{80,} 885 (2008).

\bibitem{Greiner} Greiner, M., Mandel, O., Esslinger, T., H$\ddot{a}$nsch,
T.W. \& Bloch, I. Quantum phase transition from a superfluid to a Mott
insulator in a gas of ultracold atoms. \textit{Nature} \textbf{415,} 39-44
(2002).

\bibitem{Liu-2006} Liu, W. \& Wu, C. Atomic mater of nonzero-momentum
Bose-Einstein condensation and orbital current order. \textit{Phys. Rev. A}
\textbf{74}, 013607 (2006).

\bibitem{Wu-2008} Wu, C. Orbital ordering and frustration of $p$-band Mott
insulator. \textit{Phys. Rev. Lett.} \textbf{100}, 200406 (2008).

\bibitem{Zhang-2011} Zhang, M. Hung, H.-h, Zhang, C. \& Wu, C. Quantum
anomalous Hall states in the $p$-orbital honeycomb optical lattices. \textit{%
Phys. Rev. A} \textbf{83}, 023615 (2011).

\bibitem{Xu-2013} Xu, Y., Chen, Z., Xiong, H., Liu, W. V. \& Wu, B.
Stability of $p$-orbital Bose-Einstein condensates in optical checkerboard
and square lattices. \textit{Phys. Rev. A} \textbf{87}, 013635 (2013).

\bibitem{Wirth-2011} Wirth, G., \"{O}lschl\"{a}ger, M. \& Hemmerich, A.
Evidence for orbital superfluidity in the $P$-band of a bipartite optical
square lattice. \textit{Nature Physics} \textbf{7}, 147-153 (2011).

\bibitem{Sengstock-2012} Soltan-Panahi, P., L\"{u}hmann, D., Struck, J.,
Windpassinger, P. \& Sengstock, K. Quantum phase transition to
unconventional multi-orbital superfluidity in optical lattices. \textit{%
Nature Physics} \textbf{8}, 71-75 (2012).

\bibitem{Chin} Parker, C. V., Ha, L.-C. \& Chin, C. Direct observation of
effective ferromagnetic domains of cold atoms in a shaken optical lattice.
\textit{Nature Physics} \textbf{9,} 769 (2013).

\bibitem{Zhai-1} Zheng, W., Liu, B., Miao, J., Chin, C. \& Zhai, H. Strongly
interaction effects in a superfluid ising quantum phase transition. Preprint
at http://arxiv.org/abs/1402.4569 (2014).

\bibitem{Erich} Choudhury, S. \& Mueller, J. E. Stability of a Floquet
Bose-Einstein condensate in a one-dimensional optical lattice. \textit{Phys.
Rev. A} \textbf{90}, 013621 (2014).

\bibitem{Lim-2008} Lim, L.-K, Smith, C. M. \& Hemmerich, A. Staggered-vortex
superfluid of ultracold bosons in an optical lattice. \textit{Phys. Rev.
Lett.} \textbf{100}, 130402 (2008).

\bibitem{Eckardt} Hauke, P. \textit{et al.} Non-Abelian gauge fields and
topological insulators in shaken optical lattices. \textit{Phys. Rev. Lett.}
\textbf{109,} 145301 (2012).

\bibitem{Jotzu2014} Jotzu, G. \textit{et al.} Experimental realization of
the topological Haldane model. Preprint at http://arxiv.org/abs/1406.7874
(2014).

\bibitem{FF64} Fulde, P. \& Ferrell, R. A. Superconductivity in a strong
spin-exchange field. \textit{Phys. Rev.} \textbf{135,} 550 (1964).

\bibitem{LO64} Larkin, A. I. \& Ovchinnikov, Y. N. Nonuniform state of
superconductors. \textit{Zh. Eksp. Teor. Fiz.} \textbf{47,} 1136 (1964).

\bibitem{Ian} Lin, Y.-J., Garcia, K. J. \& Spielman, I. B.
Spin-orbit-coupled Bose-Einstein condensates. \textit{Nature} \textbf{471,}
83-86 (2011).

\bibitem{Pan} Zhang, J.-Y. \textit{et al.} Collective dipole oscillation of
a spin-orbit coupled Bose-Einstein condensate. \textit{Phys. Rev. Lett.}
\textbf{109,} 115301 (2012).

\bibitem{Peter} Qu, C., Hamner, C., Gong, M., Zhang, C. \& Engels, P.
Observation of Zitterbewegung in a spin-orbit coupled Bose-Einstein
condensate. \textit{Phys. Rev. A} \textbf{88}, 021604(R) (2013).

\bibitem{Zhang} Wang, P. \textit{et al.} Spin-orbit coupled degenerate Fermi
gases. \textit{Phys. Rev. Lett.} \textbf{109,} 095301 (2012).

\bibitem{Martin} Cheuk, L. W. \textit{et al.} Spin-Injection spectroscopy of
a spin-orbit coupled Fermi gas. \textit{Phys. Rev. Lett.} \textbf{109,}
095302 (2012).

\bibitem{Kane} Hasan, M. Z. \& Kane, C. L. Topological insulators. \textit{%
Rev. Mod. Phys.} \textbf{82} 3045 (2010).

\bibitem{Qi} Qi, X. L. \& Zhang, S. C. Topological insulator and
superconductor. \textit{Rev. Mod. Phys.} \textbf{83} 1057 (2011).

\bibitem{Zhang2008} Zhang, C., Tewari, S., Lutchyn, R. M. \& Das Sarma, S. $%
p_x+ip_y$ superfluid from $s$-wave interactions of Fermionic cold atoms.
\textit{Phys. Rev. Lett.} \textbf{101}, 160401 (2008).

\bibitem{Sato} Sato, M., Takahashi, Y. \& Fujimoto, S. Non-abelian
topological order in $s$-wave superfluids of ultracold Fermionic atoms.
\textit{Phys. Rev. Lett.} \textbf{103}, 020401 (2009).

\bibitem{Fu} Fu, L. \& Kane, C. L. Superconducting proximity effect and
Majorana fermions at the surface of a topological insulator. \textit{Phys.
Rev. Lett.} \textbf{100,} 096407 (2008).

\bibitem{QiZhou} Zhang, S. \& Zhou, Q. Shaping topological properties of the
band structures in a shaken optical lattice. Preprint at
http://arXiv.org/abs/1403.0210 (2014).

\bibitem{Class1} Schnyder, A., Ryu, S., Furusaki, A. \& Ludwig, A. W. W.
Classification of topological insulators and superconductors in three
spatial dimensions. \textit{Phys. Rev. B} \textbf{78}, 195125 (2008).

\bibitem{Class2} Kitaev, A. Periodic table for topological insulators and
superconductors. \textit{AIP Conf. Proc.} \textbf{1134}, 22-30 (2009).

\bibitem{BDI-Liu} He. J. \textit{et al.} Correlated spin currents generated
by resonant-crossed Andreev reflections in topological superconductors.
\textit{Nat. Commun.} 5:3232 (2014).

\bibitem{AIII-TI-1} Liu, X.-J., Liu, Z.-X. \& Cheng, M. Manipulating
topological edge spins in one-dimensional optical lattice. \textit{Phys.
Rev. Lett.} \textbf{110,} 076401 (2013).

\bibitem{AIII-TI-2} Li, X., Zhao, E. \& Liu, W. V. Topological states in a
ladder-like optical lattice. \textit{Nat. Commun.} \textbf{4,} 1523 (2013).

\bibitem{HuiZhai} Zhang, W. \& Zhai, H. Floquet topological states in
shaking optical lattices. \textit{Phys. Rev. A} \textbf{89}, 061603(R)
(2014).

\bibitem{Konig} K$\ddot{o}$nig, M. \textit{et al.} The quantum spin Hall
effect: theory and experiment. J. Phys. Soc. Jpn.\textbf{77}, 031007 (2008).

\bibitem{LiuXJ} Liu, X.-J., Law, K. T. \& Ng, T. K. Realization of 2D
spin-orbit interaction and exotic topological orders in cold atoms. \textit{%
Phys. Rev. Lett.} \textbf{112,} 086401 (2014).

\bibitem{topoFF1} Zhang, W. \& Y, W. Topological
Fulde-Ferrel-Larkin-Ovchinnikov states in spin-orbit coupled Fermi gases.
\textit{Nat. Commun.} 4:2711 (2013).

\bibitem{topoFF2} Liu, X.-J. \& Hu, H. Topological Fulde-Ferrell superfluids
in spin-orbit-coupled atomic Fermi gases. \textit{\ Phys. Rev. A} \textbf{88,%
} 023622 (2013).

\bibitem{topoFF3} Qu, C. \textit{et al.} Topological superfluids with
finite-momentum pairing and Majorana fermions. \textit{Nat. Commun.} 4:2710
(2013).

\bibitem{topoFF4} Chen, C. Inhomogeneous topological superfluidity in
one-dimensional spin-orbit-coupled Fermi gases. \textit{Phys. Rev. Lett.}
\textbf{111,} 235302 (2013).

\bibitem{Pethick} Pethick, C.J., \& Smith, H. Bose-Einstein condensation in
dilute gases. Cambridge university press, 2002.

\bibitem{mixture} Bijlsma, M., Heringa, B. A., \& Stoof, H. T. D., \textit{%
Phys. Rev. A} \textbf{61}, 053601 (2000).

\bibitem{CJWu} Wu, C.J. \& Das Sarma, S. $p_{x,y}$-orbital couterpart of
graphene: cold atoms in the honeycomb optical lattice. \textit{Phys. Rev. B}
\textbf{77}, 235107 (2008).

\bibitem{crossover1} Ohashi, Y. \& Griffin, A. BCS-BEC crossover in a gap of
Fermi atoms with a Feshbach resonance. \textit{Phys. Rev. Lett.} \textbf{89}%
, 130402 (2002).

\bibitem{crossover2} Zhao, E. \& Paramekanti, A. BCS-BEC crossover on the
two-dimensional honeycomb lattice. \textit{Phys. Rev. Lett.}, \textbf{97},
230404 (2006).
\end{thebibliography}
\end{document}